\documentclass{pasj00}
\Received{$\langle$2010 March 4$\rangle$}
\Accepted{$\langle$2010 May 27$\rangle$}
\Published{$\langle$2010 Oct 25$\rangle$}
\SetRunningHead{Astronomical Society of Japan}{Usage of \texttt{pasj00.cls}}

\usepackage{times}

\begin{document}

\title{The Fibre Multi-Object Spectrograph (FMOS) for Subaru Telescope}

\author{
Masahiko \textsc{Kimura},\altaffilmark{1}
Toshinori \textsc{Maihara},\altaffilmark{2}
Fumihide \textsc{Iwamuro},\altaffilmark{2}
Masayuki \textsc{Akiyama},\altaffilmark{3}
Naoyuki \textsc{Tamura},\altaffilmark{1}
\\ 
Gavin B. \textsc{Dalton},\altaffilmark{5,6}
Naruhisa \textsc{Takato},\altaffilmark{1}
Philip \textsc{Tait},\altaffilmark{1}
Kouji \textsc{Ohta},\altaffilmark{2}
Shigeru \textsc{Eto},\altaffilmark{2} %
\thanks{Present address: Nikon co., Japan}
Daisaku \textsc{Mochida},\altaffilmark{2} %
\thanks{Present address: Nagoya City Science Museum, Japan}
\\ 
Brian \textsc{Elms},\altaffilmark{1}
Kaori \textsc{Kawate},\altaffilmark{2}
Tomio \textsc{Kurakami},\altaffilmark{1}
Yuuki \textsc{Moritani},\altaffilmark{2}
Junichi \textsc{Noumaru},\altaffilmark{1}
Norio \textsc{Ohshima},\altaffilmark{1}
\\ 
Masanao \textsc{Sumiyoshi},\altaffilmark{2}
Kiyoto \textsc{Yabe},\altaffilmark{2}
Jurek \textsc{Brzeski},\altaffilmark{4}
Tony \textsc{Farrell},\altaffilmark{4}
Gabriella \textsc{Frost},\altaffilmark{4}
\\ 
Peter R. \textsc{Gillingham},\altaffilmark{4}
Roger \textsc{Haynes},\altaffilmark{4}
Anna M. \textsc{Moore},\altaffilmark{4}
\thanks{Present address: Caltech Optical Observatories, 
California Institute of Technology,  
1200 E California Blvd, Pasadena, CA, USA}
Rolf \textsc{Muller},\altaffilmark{4}
Scott \textsc{Smedley},\altaffilmark{4}
Greg \textsc{Smith},\altaffilmark{4}
\\
David G. \textsc{Bonfield },\altaffilmark{5}
\thanks{Present address: University of Hertfordshire, UK}
Charles B. \textsc{Brooks},\altaffilmark{5}
Alan R.\textsc{Holmes},\altaffilmark{5}
Emma \textsc{Curtis Lake},\altaffilmark{5}
\\
Hanshin \textsc{Lee},\altaffilmark{5}
\thanks{Present address: University of Texas at Austin, TX,USA}
Ian J. \textsc{Lewis},\altaffilmark{5}
Tim R. \textsc{Froud},\altaffilmark{6}
Ian A. \textsc{Tosh},\altaffilmark{6}
Guy F. \textsc{Woodhouse},\altaffilmark{6}
\\
Colin \textsc{Blackburn},\altaffilmark{7}
Nigel \textsc{Dipper},\altaffilmark{7}
Graham \textsc{Murray},\altaffilmark{7}
Ray \textsc{Sharples},\altaffilmark{7}
and,
David J. \textsc{Robertson},\altaffilmark{7}
}

\altaffiltext{1}{Subaru Telescope, NAOJ, 650 North Aohoku Place,
 Hilo, HI, 96720,USA}
\altaffiltext{2}{Department of Astronomy, Faculty of Science,
 Kyoto University, Sakyo-ku,Kyoto,606-8502}
\altaffiltext{3}{Astronomical Institute,
 Tohoku University, Aoba-ku, Sendai, 980-8578}
\altaffiltext{4}{Anglo-Australian Observatory, Epping NSW, 1710 Australia}
\altaffiltext{5}{Department of Astrophysics, University of Oxford,
 Keble Road, Oxford OX1 3RH, UK}
\altaffiltext{6}{STFC Rutherford Appleton Laboratory, Chilton, Didcot,
 Oxfordshire, OX11 0QX, UK}
\altaffiltext{7}{Department of Physics, University of Durham,
 South Road, Durham DH1 3LE, UK}

\email{mkimura@subaru.naoj.org}


\maketitle

\begin{abstract}
Fibre Multi-Object Spectrograph (FMOS) is the first near-infrared 
instrument with a wide field of view capable of acquiring spectra 
simultaneously from up to 400 objects. 
It has been developed as a common-use instrument for the F/2 prime-focus 
of the Subaru Telescope. The field coverage of 30$^\prime$ diameter 
is achieved using a new 3-element corrector optimized in the near-infrared 
(0.9-1.8$\mu$m) wavelength range. Due to limited space at the prime-focus, 
we have had to develop a novel fibre positioner called 
''Echidna'' together with two OH-airglow suppressed spectrographs. 
FMOS consists of three subsystems: the prime focus unit 
for IR, the fibre positioning system/connector units, 
and the two spectrographs. After full systems integration, FMOS was installed 
on the telescope in late 2007. Many aspects of performance 
were checked through various test and engineering observations. 
In this paper, we present the optical and mechanical components 
of FMOS and show the results of our on-sky engineering observations to date.
\end{abstract}

\section{Introduction}
In recent years, multi-object spectroscopy surveys on large telescopes have
proven to be an essential technique to investigate galaxy formation
and other statistical parameters of the Universe. Consequently, most 
large telescopes now have some kind of multi-object spectrograph. 
Statistical observations using near-infrared 
spectrographs have played an important role in the study of 
galaxy evolution for years. This is largely because the results can be 
directly compared with studies of local galaxies at optical wavelength. 
These observations, however, require a considerable amount of telescope time. 
The multi-object spectrograph in the near-infrared (NIR) wavelength 
range is the key instrument to make such a study usual. 

Using the MOIRCS (multi-object NIR spectrograph) 
instrument (\cite{suzuki2009}) on the Subaru telescope,
it is already possible to obtain spectra from 30-50 objects 
within a $4'\times 7'$ field of view using cold slit masks. The FMOS
instrument however, is capable of acquiring $\sim$ 10 times 
more objects within $\sim$ 25 times wider field of view. Such a large 
increase in objects and wide field of view is useful not only for the study 
of the galaxy evolution and variation with galaxy environment, 
but also for investigating star-forming regions, cluster formation, 
cosmology, and so on. 

\section{Design of the instrument}
\subsection{Overview}
\begin{figure*}[thbp]
  \begin{center}
    \FigureFile(135mm,105mm){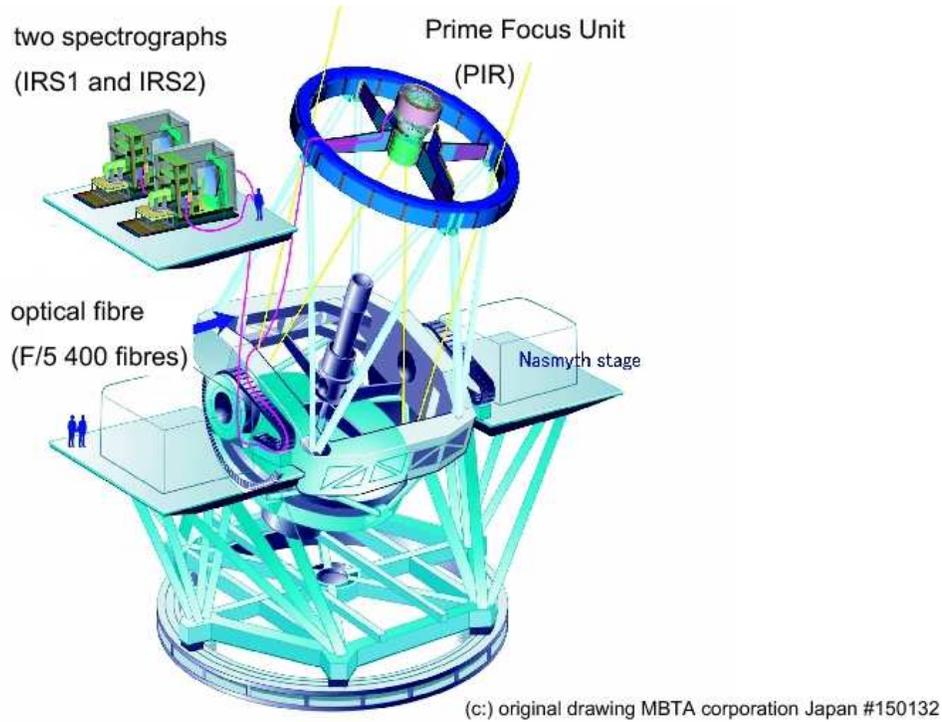}
  \end{center}
  \caption{The mechanical layout of FMOS on Subaru telescope, 
showing the locations of the prime focus unit (PIR), fibre bundle, 
and the two spectrographs (IRS1 and IRS2).}\label{fig:subaru}
\end{figure*}

FMOS is a fibre-fed NIR spectrograph of Subaru Telescope 
(\cite{kimura2003};\,\cite{maihara2000}). 
FMOS has a capability to acquire spectra from 400 targets simultaneously 
in a 0.9-1.8 
$\mu$m wavelength range with a field coverage of 30$^\prime$ 
diameter at the prime-focus of the telescope 
(figure \ref{fig:subaru}). 
FMOS consists of three subsystems; 1) the Prime focus unit for 
InfraRed (PIR), 2) the fibre positioning system/connector units called 
Echidna, 3) the two cooled infrared spectrographs (IRS1 and IRS2). 

The PIR containing Echidna fibre positioner is mounted 
on the centre of the top-ring of the telescope.
It has an attachment mechanism, an optical alignment mechanism, 
and corrector lens system (figure \ref{fig:pir1}).

Echidna is a novel fibre positioning system collecting light from
the 15cm diameter focal plane at the prime-focus of the Subaru Telescope.
Tilting fibres are attached to a carbon-fibre ''spine'' and are
moved using a quadrant tube piezo actuator 
(\cite{brzeski2004};\,\cite{gillingham2000},\,2003;\,
\cite{moore2003}). 
In front of the array of fibres, the Focal Plane Imager (FPI) 
is located on the XY gantry. The FPI is a dual imaging system 
capable of measuring the fibre 
positions for the instrument calibration as well as acquiring 
sky images for field acquisition. 
Using several iterations with FPI, 
Echidna completes a field reconfiguration in less than 15 minutes. 
The 400 science fibres are divided equally and fed to two cooled spectrographs 
installed on the fourth floor of the telescope dome
(\cite{murray2003},\,2004,\,2008). 
The interconnecting optical feed is a fibre-optic 
downlink with a length of $\sim$ 70 m.

The two cooled spectrographs have the capability of eliminating 
strong OH-airglow lines in the $J$- and $H$-band, which are 
the major sources of background radiation at NIR wavelength
(\cite{dalton2006},\,2008;\,\cite{iwamuro2006},\,2008;\,
\cite{lewis2003},\,2004). 
The suppression system provides a substantial S/N gain for
observations at low spectral resolution, thus allowing large
simultaneous spectral coverage from a single 2k $\times$ 2k detector 
 (\cite{iwamuro1994},\,2001;\, \cite{maihara1993},\,1994). 
To reduce the thermal background in the spectrographs, 
all of the optics are assembled in a large refrigerator 
and cooled to below $-$50$^\circ$C. Each spectrograph has two modes 
of spectral resolutions: a high-resolution mode 
with $\lambda/\Delta\lambda=2200$ and a low-resolution mode 
with $\lambda/\Delta\lambda=500$. 
The low-resolution mode is realized by using a VPH 
(Volume Phase Holographic) grating as an anti-disperser. 
The NIR camera uses a HAWAII-2 HgCdTe detector. It
covers the entire $J$- and $H$-bands in low-resolution mode. 

The conceptual design of FMOS started in 1998, and all the mechanical 
components were assembled and mounted on the telescope in late 2007.
Intensive commissioning observations started in December 2007
and first light was accomplished in May 2008. 
FMOS is expected to explore various scientific frontiers 
from nearby substellar objects to the large scale structure
of the distant universe.

\subsection{PIR: Prime focus unit for IR}
\subsubsection{Interface mechanism}
\begin{figure*}[htbp]
  \begin{center}
   \FigureFile(80mm,80mm){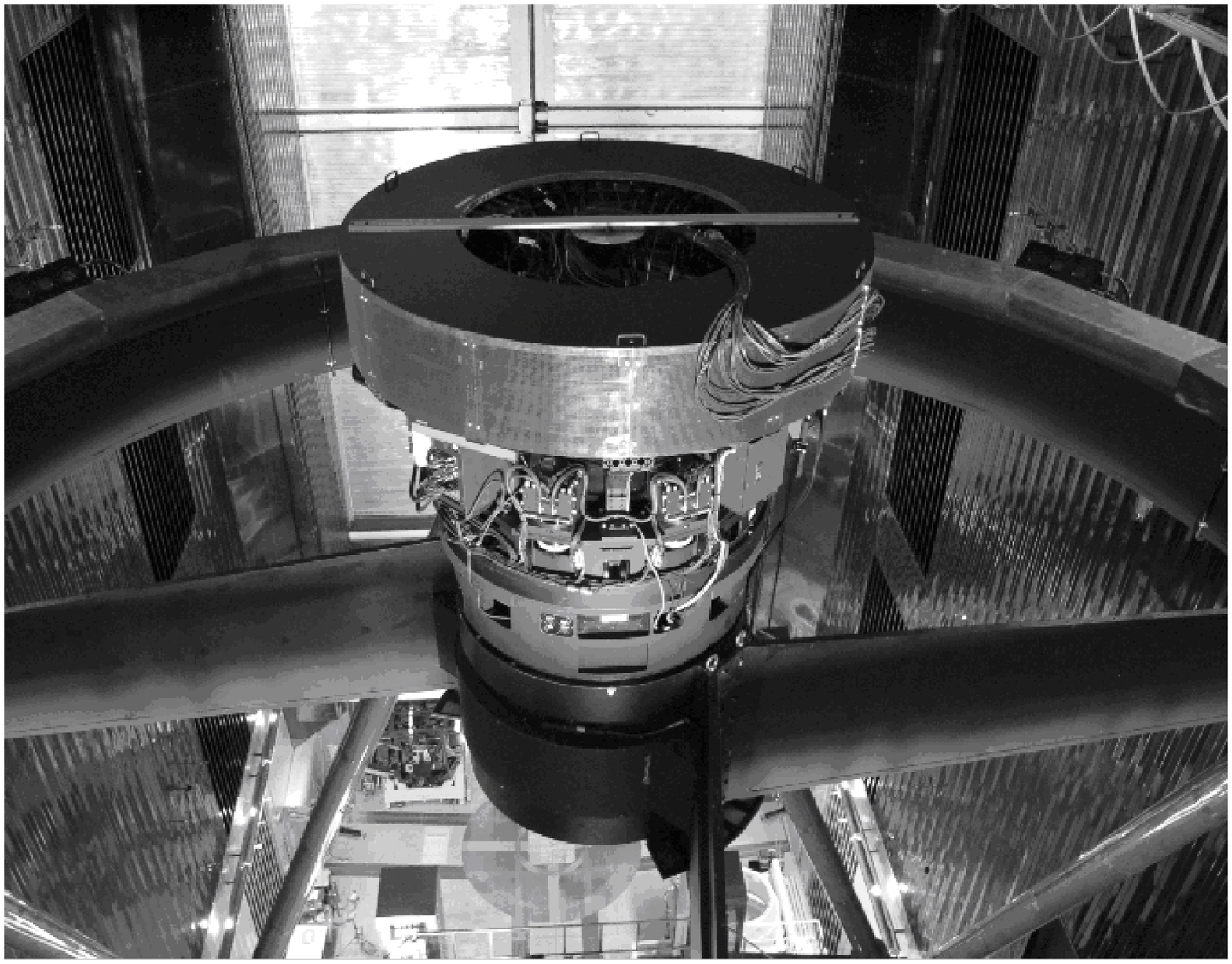}
   \quad
   \FigureFile(60mm,80mm){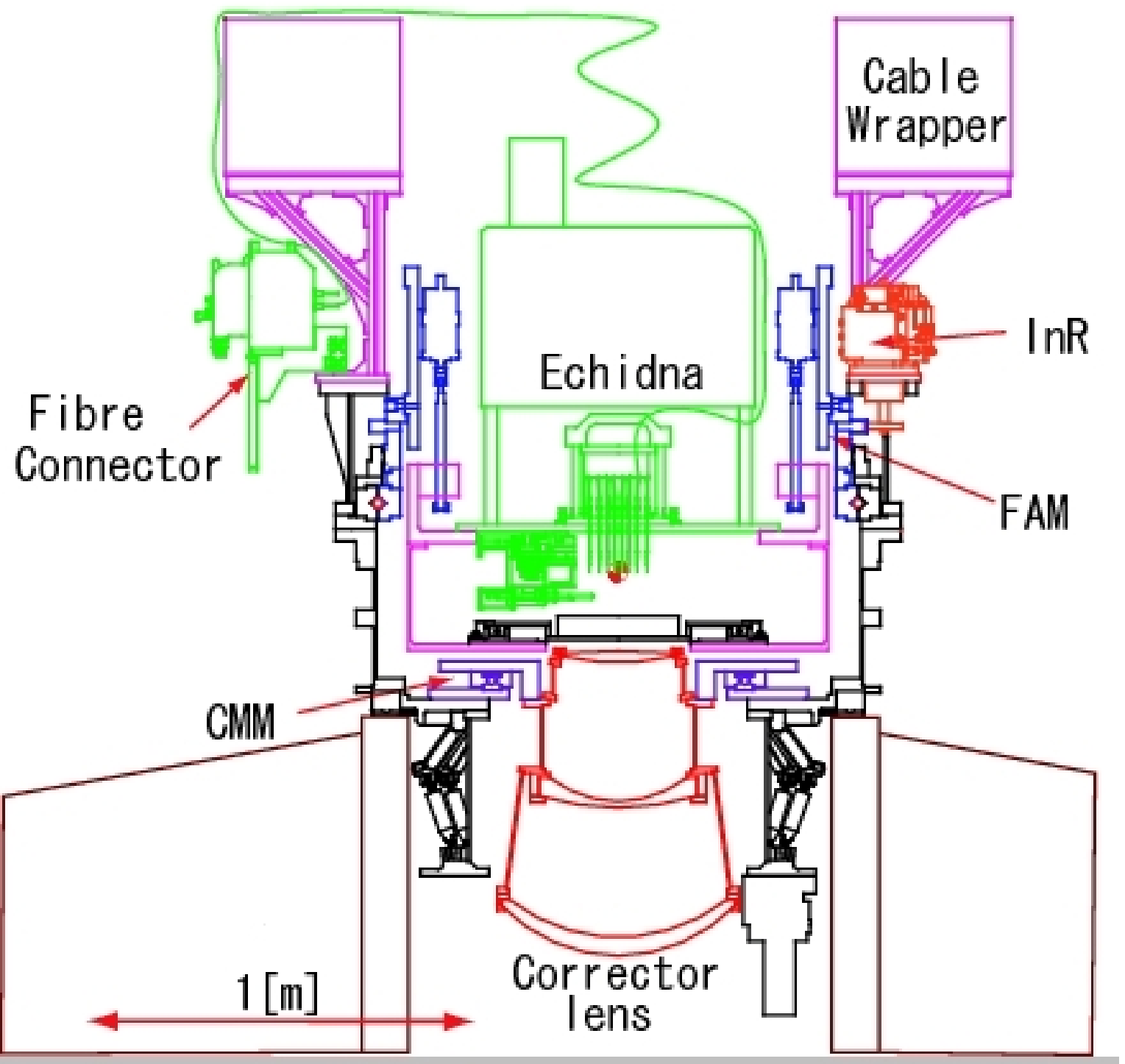}
  \end{center}
  \caption{The prime focus unit attached to Subaru
 Telescope(left). Sectional drawing of PIR(right).}\label{fig:pir1}
\end{figure*}

Subaru Telescope has two prime focus units and three secondary 
mirrors exchangeable semi-automatically in about six hours in the daytime. 
One prime focus unit is optimized for a visible light 
and the other (i.e. PIR) is for NIR wavelength region. The PIR is developed 
as the front-end unit for FMOS, attached to the centre of the top ring. 
The Echidna fibre positioner with FPI, 
and the Shack-Hartmann camera are 
installed in the instrument bay, which is supported by the 
outer shell structure with the Focus-Adjustment Mechanism 
(FAM) and the instrument rotator (figure \ref{fig:pir1}). 
The corrector lens system is fixed on the bottom of the outer 
shell structure via the Corrector-Movement Mechanism (CMM) to 
compensate for the offset of the corrector lens system from 
the optical axis of the primary mirror 
with rotation/tilt of the telescope. 
A computer, cable wrapping system and the 
fibre connector are also attached to this outer shell structure.

We can make an optical alignment between the corrector axis and 
the telescope axis by CMM and FAM using a defocused star image taken 
by the sky camera of FPI (see 2.3.1). 
We can also measure the optical aberration of the telescope 
including the corrector lenses in detail using the Shack-Hartmann camera 
at the centre of the field. 
The results of Shack-Hartmann camera are analysed in a series 
of Zernike's coefficient 
by the mirror analysis software of Subaru telescope. 
On the basis of these measurement and corrections, 
the best XY position of CMM and the best Z position of FAM 
can be determined as a function of the telescope elevation-angle. 

\subsubsection{Prime focus corrector}
Faint galaxies are considered as main targets of FMOS. 
The measured intrinsic half-light radius 
of the faint galaxy is typically 0$.\!\!^{\prime\prime}$3 
at $H$=20.5 (\cite{yan1998}).
The typical atmospheric seeing 
is FWHM=0$.\!\!^{\prime\prime}$6 on the summit of Mauna Kea, 
yields about 0$.\!\!^{\prime\prime}$9 diameter 
for an 80 $\%$ energy circle of the faint galaxies. 
Thus, we select the aperture diameter of the fibre core to 
1$.\!\!^{\prime\prime}$2 on the sky.
This aperture diameter corresponds to about 100 $\mu$m 
on the F/2.1 focal plane of Subaru telescope.
 
A telecentric focal surface has been chosen
to decrease the misalignment between an object
and its fibre aperture over the whole field.
The three (BSM51Y) spherical lenses of the corrector system allow 
acquisition over its 30$^\prime$  diameter field of 
view in the NIR wavelength range. 
The first corrector element has a diameter of 590mm. 

\begin{figure}[htbp]
  \begin{center}
    \FigureFile(80mm,60mm){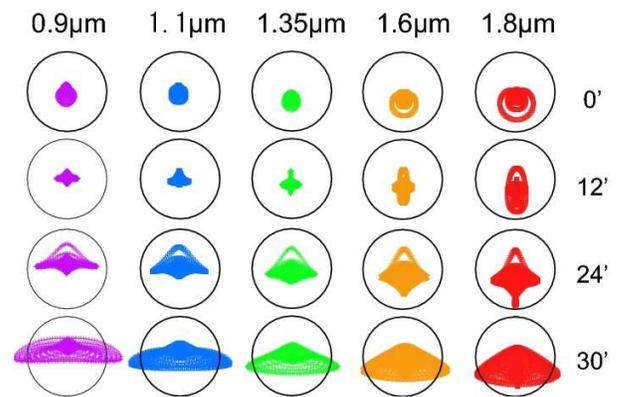}
   \end{center}
  \caption{Spot diagram of the corrector at zenith position. 
Spots at the wavelengths of 0.9, 1.1, 1.35, 1.6 and 1.8 $\mu$m 
vs. the positions of 0$^{\prime}$(centre), 6$^{\prime}$, 12$^{\prime}$ 
and 15$^{\prime}$(edge) are plotted.
The circle represents the fibre diameter of 100 $\mu$m.
}\label{fig:focus1}
\end{figure}

The optical aberration of this corrector system is less than 
0$.\!\!^{\prime\prime}$2 (15$\mu$m) in FWHM, small enough 
compared to the fibre core diameter of 1$.\!\!^{\prime\prime}$24 
(100$\mu$m). 
The 80$\%$ encircled energy diameter is 0$.\!\!^{\prime\prime}$7 
at the edge of the field of view including the chromatic aberration 
of magnification (figure \ref{fig:focus1}). 
In the NIR wavelength range of 0.9-1.8$\mu$m, the atmospheric 
dispersion is 0$.\!\!^{\prime\prime}$1 and 0$.\!\!^{\prime\prime}$3 
at zenith distance of 30 and 60 degrees, respectively. 
Since these dispersion displacements are considerably smaller 
than the diameter of the 
science fibre, we do not use an Atmospheric Dispersion Corrector (ADC).

\subsection{Fibre positioning system}
\subsubsection{Fibre positioner and focal plane imager}
The prime focus of Subaru telescope has a fast F-ratio and 
a wide field of view, however it is too small (about 150mm 
in diameter) to use a fibre positioner with magnetic buttons 
such as 2dF/AAT (\cite{lewis2002}). 
Therefore we have developed a novel fibre positioning system, 
Echidna, capable of positioning up to 400 fibres at the prime 
focus (figure \ref{fig:pir1}). 
It is based on the use of piezo tube actuators tilting 
the fibres by electric pulse trains of roughly saw-tooth shape. 
Each science fibre has a length of $\sim$ 160mm from the tip of the fibre 
to the pivot ball on the tube piezo actuator, covering 
a circular area of $\sim$ 7mm in radius by tilting up to 2.5 degrees. 
The optical loss caused by the tilt to the incident beam 
is less than 10 $\%$ even at maximum tilt.
To limit coupling efficiency losses
between a target and a fibre, 
the fibre is positioned within 10 $\mu$m of its target.

\begin{figure}[htbp]
  \begin{center}
    \FigureFile(80mm,95mm){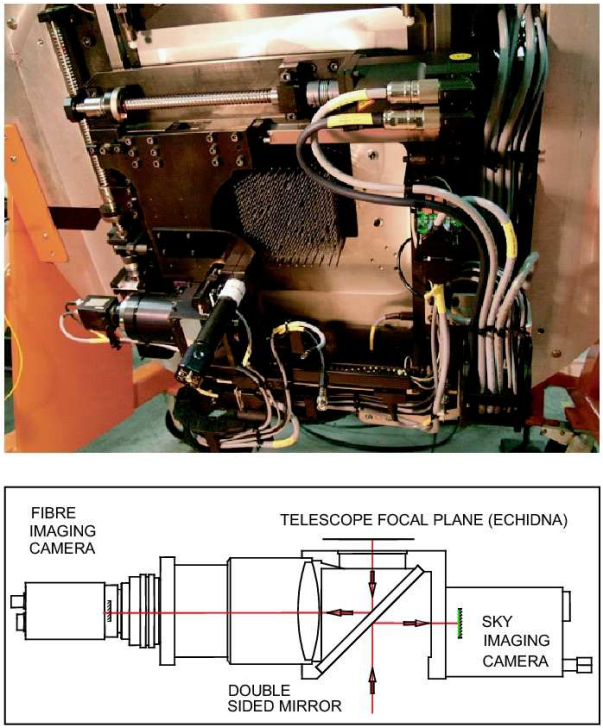}
  \end{center}
  \caption{Echidna fibre positioner with FPI. 
The fibre camera is used for the fine alignment 
of the science fibres as well as the guide-fibre bundles, 
while the sky camera is used for the correction of the initial pointing 
error of the telescope.
The whole view of Echidna, focal plane (top), design of the
 FPI (bottom).
}\label{fig:echi}
\end{figure}

In order to achieve the required positioning accuracy, 
we utilise the FPI for spine position feedback (figure \ref{fig:echi}).
The FPI has two cameras; the sky camera and the fibre camera, 
mounted on the XY stage. 

The sky camera checks the telescope pointing 
by measuring the positions of several bright stars 
($R \le$ 15 AB mag).
The fibre camera, which divides the focal plane into 59 small sub-fields, 
checks each fibre's position 
using a back-illumination system.
After measuring the positioning error of each fibre from 
the image taken by the fibre camera, the positioner moves 
each spine (in parallel) to correct for the error. 
To achieve a desired positioning accuracy of less than 10$\mu$m, this
process is repeated $\sim$ 7 times.

\begin{figure}[htbp]
  \begin{center}
    \FigureFile(80mm,60mm){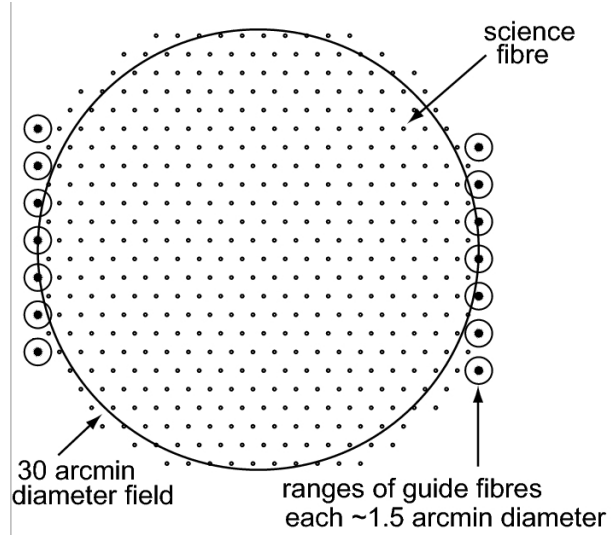}
  \end{center}
  \caption{Layout of 400 science fibres (at home positions) and 
14 guide-fibre bundles in focal plane.
Circles correspond to the patrol area of the guide-fibre bundles.
The guide bundles are attached to spines, positioned by tilting 
in the same way as science fibres. 
The fibres in the guide-fibre bundle are closely packed at their input end, 
while they are separated at the output end to enable an independent 
intensity measurement by the guiding camera in Echidna.
}\label{fig:fov}
\end{figure}

Echidna has 14 guide-fibre bundles 
for auto guiding - they are located on each side of the field.
Each guide-fibre bundle consists of 7$\times$50 $\mu$m diameter fibres. 
These bundles are imaged onto a cooled CCD camera through OG570
filter, where stars with $R$=16.5 mag. or brighter 
are usable for auto guiding.

\subsubsection{Fibre cable and fibre connector}
The fibre connector is required to allow the PIR
to be removed from the telescope and replaced with
other instruments.
The connector box is attached to the side surface of the PIR, 
while the conjugated plug is usually parked at the junction 
between the centre of the top ring and the spiders.
Additionally, 
the connector has two more roles apart from detaching the PIR unit; 
1) conversion of F-ratio from F/2 fibres of Echidna into F/5 
fibres of the bundle fed by IRS1 and IRS2, 2) back-illumination of F/2 fibres to flash their tips 
at the prime focus during the exposure of the fibre camera 
(\cite{murray2003},\,2004). 
The incident light at the focal plane has a focal ratio of F/2, 
which is not optimal when light is to be transmitted through a long
fibre. Furthermore, at the spectrograph,  
such a divergent exit beam would present significant challenges 
for the design of the spectrograph collimator optics.
The expected efficiency of a plano-convex lens coupling 
in the connector is more than 95$\%$ over the wavelength range. 
The F/2 fibres are 7.6 m long, to cover the routing distance from  
the focal plane of Echidna to the site of connector.
The F/5 fibres are 62 m long, from connector to the spectrographs.

The back-illumination has a movable coupling prism and
LEDs which is necessary to measure the exact position of 
each spine with each exposure on sky.

\subsection{Cooled spectrograph}
\subsubsection{Overview}
FMOS has two spectrographs, IRS1 was developed by Kyoto University 
and IRS2 was developed by University of Oxford and Rutherford Appleton
Laboratory, to obtain 2$\times$200 spectra simultaneously in the NIR 
wavelength range (\cite{dalton2006},\,
\cite{iwamuro2006},\,\cite{kimura2003},\,\cite{tosh2004}). 
Although the mechanical components are completely different, 
all the optical components and parameters of these spectrographs 
are identical. Each spectrograph has two spectral resolution modes; 
the low-resolution mode covers all the wavelength range of 
0.9 - 1.8$\mu$m with one exposure, while the high-resolution mode 
requires four exposures at different camera positions to cover the 
full wavelength range.

\begin{figure}[htbp]
  \begin{center}
    \FigureFile(80mm,60mm){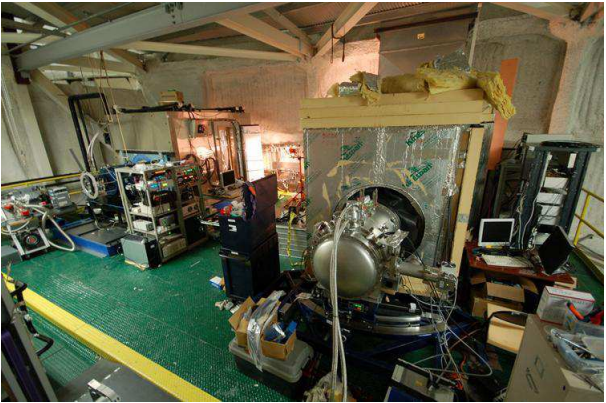}
  \end{center}
  \caption{Two cooled NIR spectrographs. Nearest is IRS2 -
the other is IRS1.}\label{fig:spect}
\end{figure}

The sky background in the NIR region is dominated by many narrow 
emission lines by OH-airglow. 
During the night, the OH-airglow lines vary in brightness 
by 5-10$\%$ on a timescale of 5-15 minutes 
as atmospheric wave flow changes the column density of 
excited molecules in the line of sight. 
These strong lines affect not only the background noise 
at the specified wavelength, but also the stability 
of the detector itself. 
These bright OH-airglow lines are suppressed by the mask 
mirror in the spectrograph.
In addition, thermal emission from the optical components 
is also a considerable noise source. 
There is no cold stop in the spectrograph: 
all optical components of spectrograph are cooled in a large refrigerator 
to keep the temperature below $-$50 $^\circ$C (figure \ref{fig:spect}).

\subsubsection{Optics of spectrograph}
\begin{figure}[htbp]
  \begin{center}
    \FigureFile(80mm,60mm){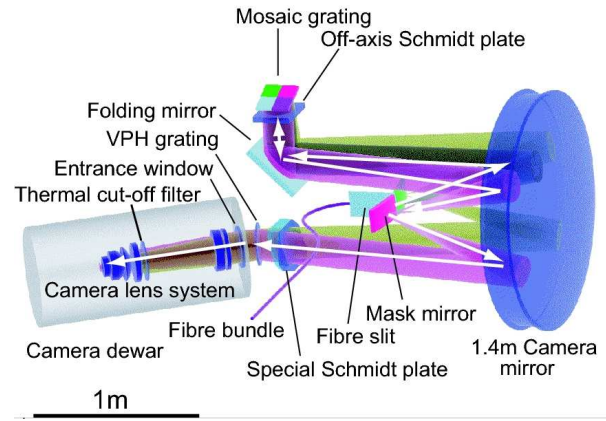}
  \end{center}
  \caption{Optical design of the spectrograph. 
Optical system from the incident fibre slit to the mask mirror 
(upper half of the drawing) is the double-path Schmidt system. 
The reflective mosaic grating and the VPH  grating are located 
at the primary and the secondary pupil, while the primary and 
the secondary spectra are focused onto the mask mirror and the detector
respectively.
}\label{fig:opt1}
\end{figure}

The CIRPASS (Cambridge InfraRed Panoramic Survey Spectrograph) instrument, 
is a fibre-fed and multi-object NIR spectrograph, 
which has been commissioned by Ian Parry et al (\cite{parry2004}).
In the initial design stage, we adopted the optical layout 
for OH suppression similar to the CIRPASS instrument.
Figure \ref{fig:opt1} shows the optical layout of the spectrograph. 
These spectrographs are designed to have a capability of OH-airglow 
suppression by focusing the primary spectra onto the mask mirror 
using double-path Schmidt optics. 
The secondary spectra are obtained by the 2048$\times$2048 infrared array 
detector HAWAII-2 installed in the camera dewar. 
The length of the incident fibre slit is 12cm, which is reduced 
to $\sim$ 4cm on the detector. 

\begin{figure}[htbp]
  \begin{center}
    \FigureFile(80mm,60mm){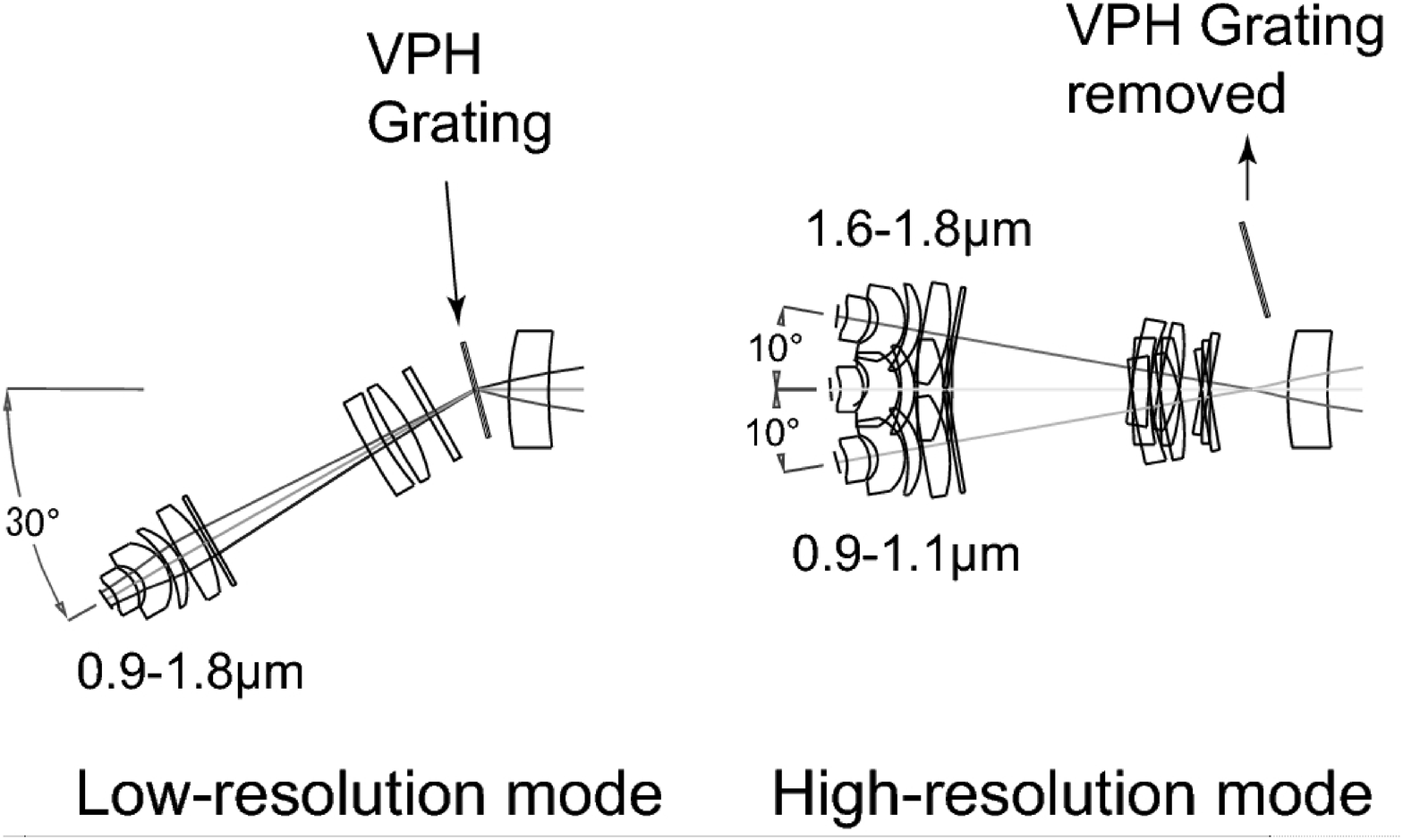}
  \end{center}
  \caption{The position of the camera dewar in each spectral mode.
}\label{fig:opt2}
\end{figure}

Although the typical spectral resolution of the primary spectra 
is $\lambda/\Delta\lambda=2200$, we can choose the spectral 
resolution of the secondary spectra $\lambda/\Delta\lambda=500$ 
in low-resolution mode ($\lambda/\Delta\lambda=2200$ 
in high-resolution mode with (without) the VPH grating,
respectively, into the optical path at secondary 
pupil position to reduce the dispersion given by the first grating. 
Spectra with the full wavelength range (0.9-1.8$\mu$m) are 
obtained by a single exposure in the low-resolution mode, 
while four exposures at different camera positions are required
to cover the same wavelength range in the high-resolution mode 
(figure \ref{fig:opt2}). 

Each fibre slit contains 200 fibres with a centre-to-centre spacing 
of 600$\mu$m (almost twice the 280$\mu$m fibre diameter), 
which consists of 10 subsets of 20 fibres separated with an additional small spacing
between them. The fibres in this fibre slit are arranged in a fan 
shape along the curvature of the focal sphere of the Schmidt optical
system. 
The output beam is nominally F/5, with up to 10 $\%$ of the light
emerging into F/4.2 which is controlled back to F/4.7 using field lenses
on the output of the slit blocks (\cite{murray2003}).

The spherical camera mirror is a honeycomb light-weight mirror with 
a diameter of 1.4m and a weight of 137kg. 
The radius of curvature is 1940mm. 
The slit is placed 951 mm in front of the collimator to produce a
single beam diameter of $\sim$200mm. 
The reflective surface is silver with sapphire over-coating.

The off-axis Schmidt plate with dimensions of 280mm $\times$ 236mm 
($\times$ 33mm in thickness) is manufactured using a high-accuracy 
grinding process (without polishing). 
The accuracy of the total shape is 0.3$\mu$m and the local roughness 
is 80nm rms. 
The material is fused silica with anti-reflection coatings on both sides.

The grating at the primary pupil position is composed of 2$\times$2 mosaic 
of usual reflective gratings with a groove density 
of 500 g/mm blazed at 1.35$\mu$m for the incident angle of 20$^\circ$,
because the diffraction limited performance is not
required and the cost reduced considerably. 
This 2$\times$2 mosaic grating is controlled by using 8 pieces of picomotor
actuators for adjustment the direction of the four gratings under
refrigerated condition.
The total dimensions of this grating is 230mm$\times$210mm.

\begin{figure}[htbp]
  \begin{center}
    \FigureFile(75mm,90mm){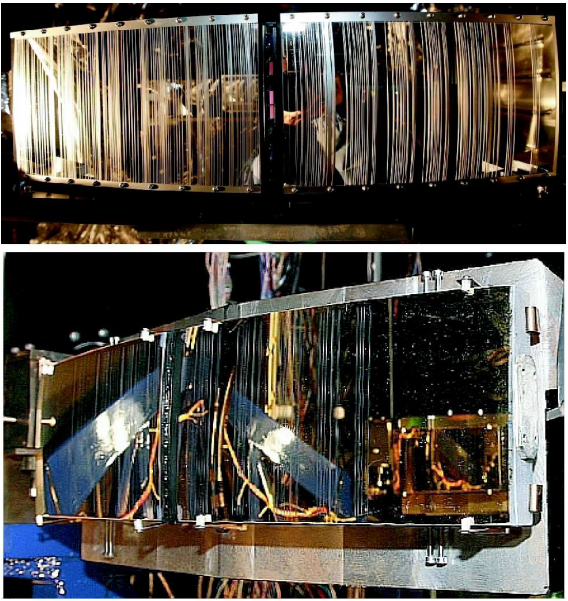}
  \end{center}
  \caption{Mask mirror in IRS1(top) and IRS2(bottom). 
These masks have patterns of the airglow lines 
in the $J$-band and $H$-band. 
The airglow mask of IRS1 is made of a thin (0.2mm 
in thickness) stainless-steel plate 
at the positions of the strong 
OH-airglow lines, and blackened to absorb the OH light 
(\cite{iwamuro2006}). 
The airglow mask of IRS2 is printed directly on the surface 
of the mirror by photochemically etching away the reflective 
gold coating of the mirror so that OH light 
is absorbed in the substrate and mount of the mirror 
(\cite{lewis2004}).
}\label{fig:mask}
\end{figure}

The mask mirror consists of the two spherical-convex mirrors 
with a curvature radius of 993mm(figure \ref{fig:mask}). 
Each mirror has dimensions 
of 280mm$\times$140mm, arranged at both sides of the entrance 
fibre slit. The gap between these two mirrors corresponds to the
atmospheric absorption band at $\sim$1.4$\mu$m 
in the primary spectra focused onto these mirrors. 
The methods of masking strong OH-airglow lines are different 
between IRS1 and IRS2. 
The airglow mask of IRS1 is made of a thin (0.2mm 
in thickness) stainless-steel plate which is processed 
by photochemical etching to leave material only 
at the positions of the strong OH-airglow lines, 
and blackened to absorb the OH light 
(\cite{iwamuro2006}). 
On the other hand, the airglow mask of IRS2 
is printed directly on the surface of the mirror 
by photochemically etching away the reflective 
gold coating of the mirror so that OH light 
is absorbed in the substrate and mount of the mirror 
(\cite{lewis2004}). 
A total of 283 OH and O$_2$ air-glow lines 
are rejected on the mask mirror, 
thus the effective opening area is 78.8 $\%$. 
The width of the each mask element is 0.4 mm, which is a factor
1.4 times the width of the fibre core diameter of 280 $\mu m$ on the
fibre slit in IRS1. 
The suppression factor of the OH-airglow line 
is about 10 and  the estimated gain is 1 mag, 
if observations are sky background limited.

The special Schmidt plate has two aspherical concave/convex 
surfaces with dimensions of 315mm$\times$251mm 
($\times \sim$90mm in thickness), 
which are achieved using a combination of two elements having 
concave-flat and flat-convex shape (figure \ref{fig:opt1}). 
Both aspherical surfaces are processed by ''ELID'' 
(ELectrolytic In-process Dressing) 
grinding method. The accuracy of the total shape is 3$\mu$m 
and the local roughness is 100nm rms. 
The material is fused silica with an anti-reflection 
coating on both sides of each element.

The VPH grating with a diameter of 262mm$\phi$ and 10mm in thickness 
is used at the secondary pupil to reduce the dispersion power 
in the low-resolution mode. 
The line density is 385 g/mm and the peak of
diffraction efficiency is around 1.3 $\mu m$ 
at the Bragg condition (\cite{tamura2003}).  
The material of the substrate is Starphire glass (similar BK7) 
with anti-reflection coatings on both sides.
No significant deterioration has been observed after a number of heating and cooling cycles.
Additionally the diffraction efficiency is nearly independent 
of temperature between 200 and 280 K. 

The camera lens system in the dewar consists of 6 spherical 
lenses (including one aspheric surface), an entrance window,
and the thermal cut-off filter with the cut-off wavelength of
1.8$\mu$m. 
The material of all the camera elements is fused silica with 
the maximum diameter of 250mm. The first order of the axial 
chromatic aberration of this system is corrected by tilting 
the detector along the dispersion axis. 
The focus position and this tilt angle are adjusted to the best position 
of the selected observation mode.

\begin{figure}[htbp]
  \begin{center}
    \FigureFile(70mm,60mm){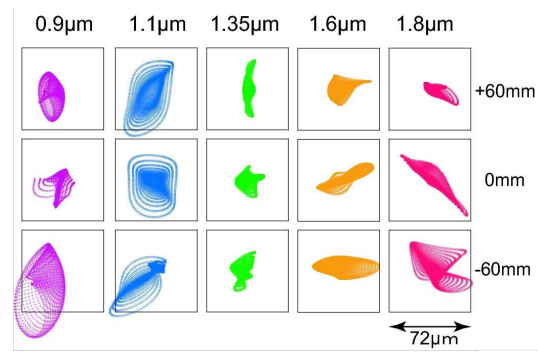}
  \end{center}
  \caption{Spot diagram on the detector in the low-resolution mode. 
Spots at the wavelengths of 0.9, 1.1, 1.35, 1.6 and 1.8 $\mu$m 
vs. the positions of +60mm(top of the fibre slit), 0mm, -60mm(bottom of the fibre slit) 
are plotted. The box represents 4 pixels of the detector.}\label{fig:fspot}
\end{figure}

The designed spot images in the low-resolution mode are shown 
in figure \ref{fig:fspot}. 
The 80$\%$ encircled energy diameter is less than 50$\mu$m 
(2.5 pixels) in this mode as well as in the high-resolution mode,
smaller than the real image size of a single fibre core 
of 80$\mu$m (4 pixels). 

\subsection{Control system}
FMOS control system is designed as a network of subsystem controller PCs. 
The central PC called FMOS OBCP (OBservation Control Processor)
interfaces with the telescope computer system, and also
supervises all the FMOS subsystems through the network. 
The OBCP links to two PIR-dedicated PCs, IRS-dedicated 
PCs, and Subaru computers. 
One PIR-PC controls FAM, CMM, and the stage of the Shack-Hartmann 
system. Another PIR-PC takes part in controlling Echidna positioner 
as well as the auto guiding signal and FPI system. 
The IRS-PCs control all the parts of the spectrograph: 
many cooled optical stages and the dewar stage, 
the cooling system of refrigerator, and the data acquisition and
transfer (\cite{eto2004},\,\cite{dipper2004}).

We use two HAWAII-2 array detectors for IRS1 and IRS2 developed by
different institutes. 
The readout electronics of IRS2 camera is a SDSU IR controller system 
(\cite{leach2000}) using the software interface developed for WFCam by
the UKATC (\cite{hirst2006}), 
while that for IRS1 is newly developed one, which adopts the similar
front-end circuit of MOIRCS (\cite{ichikawa2003}) 
combined with the Messia-V system (\cite{nakaya2004}). 
The Messia-V provides a package of the "Clock Sequencer" and the "Frame
Grabber", which has been incorporated in the CCD data acquisition system of 
the optical instruments of Subaru telescope.

\section{Performance}
\subsection{Engineering observation}
The conceptual design started in 1998, and all the mechanical 
components were assembled and mounted on the telescope in late 2007.
Since May 2008, engineering observations have been carried 
out several times to check the basic performance of FMOS. 
After measurement of the image quality at the prime focus, 
the total system efficiency from the corrector lenses to 
the detector was estimated from the results of these observations 
as well as from the test using a black body source. 
The positioning accuracy of Echidna was calibrated using several observations 
of a part of the galactic plane or open clusters under various
conditions. 
The guiding stability using the guide-fibre bundles was 
also examined from the dispersion of the error signal during the observations. 

\subsection{Optical performance of the PIR}
In the first stage of our engineering run, 
we made a rough adjustment of the corrector axis (CMM axis) 
and the telescope optical axis 
using a defocused bright-star image taken 
by the sky camera of FPI. 
The position of the focus is adjusted by FAM 
using a star image taken by the same camera. 
The typical FWHM of the point source 
is 0$.\!\!^{\prime\prime}$6 in the optical wavelength.

After the rough positioning, 
we measured the optical aberration of the telescope 
including the corrector lenses in detail using the Shack-Hartmann camera 
at the centre of the field. In this camera, the wave-front error 
was measured as the position shifts of multiple-images of a single star 
divided by a micro lens array at the pupil position. 
The results were expanded in a series of Zernike's coefficient 
by the mirror analysis software of Subaru telescope. 
On the basis of these measurement and corrections, 
the best XY position of CMM could be determined 
as a function of the telescope elevation angle. 
The Coma aberration term in the calculated Zernike's 
coefficients is sensitive to the XY position of CMM, 
while defocus term corresponds to the position of FAM.
We found that the active control of CMM as a function of the elevation 
angle of the telescope is not necessary at the elevation angle larger 
than 40$^\circ$.
The aberration coefficient indicates that the spherical 
and the coma aberration is small enough, 
while the astigmatism component contributes 
to the image size of $\sim$ 0$.\!\!^{\prime\prime}$4 even 
at the centre of the field (\cite{kimura2008}). 

The tilt and distortion of the focal plane is also checked 
by the sky camera.
We observed the field including an open cluster with tiling the sky
camera to determine the direction of the XY axes of FPI coordinate, 
the pixel scale, and the distortion map of the field. At the edge of the field 
(15$^\prime$ off-centre position), the position of the star shifts 
about 10$^{\prime\prime}$ outward because of the distortion. 
We can identify the optical axis of the telescope including 
the corrector lenses as the centre position of this distortion pattern. 
As a results of the distortion measurement, we found the position difference 
between the rotator axis and the distortion centre is 0$^\prime$.89
(4.3mm) - in other words, 
the distortion pattern moves along a circle during observation. 
This movement causes 0$.\!\!^{\prime\prime}$24 shift at the outer edge 
region of the field for 13$^\circ$ rotation of the instrument rotator. 
Since neither the optical axis of the telescope nor the rotator axis 
can be adjustable, we have to readjust (''tweak'') a part of the outer fibres 
blindly between long exposures. 
No tilt of the focal plane was detected in this measurement.

\subsection{Positioning accuracy of Echidna}
We measured the displacements of the tips of the fibres whilst 
changing the elevation angle from zenith to 30 degrees. 
Almost all the measured displacements are distributed within 
the range of 50$\pm$10$\mu$m including the displacements 
of the guide-fibre bundles. Here, the average offset of 50$\mu$m 
has no effect on the positioning accuracy, because the 
average flexure of the fibres will be cancelled 
by auto-guiding the telescope using the guide-fibre bundles. 
The measured scatter is sufficiently small to keep the targets 
within 10$\mu$m (0$.\!\!^{\prime\prime}$12) accuracy during a long exposure. 
Both the stability of the fibre positions in two hours 
and the measurement accuracy of the fibre camera in FPI 
are $\sim$2$\mu$m, negligible compared with the flexure 
variation of the fibres. 
Sufficient configuration accuracy can be achieved with 
7 iterations of readjustment using FPI (\cite{akiyama2008}): 
on average 98$\%$ of the fibres
reach target positions within 10$\mu$m 
at above zenith distance of 60 degree. 
In each configuration, about 10 fibres cannot reach 
the target positions due to collisions etc. 
Since 100 seconds are needed to measure the positions 
of 400 fibres with the FPI, 
Echidna takes about 15 minutes to complete 7 iterations. 
\begin{figure}[htbp]
  \begin{center}
    \FigureFile(75mm,75mm){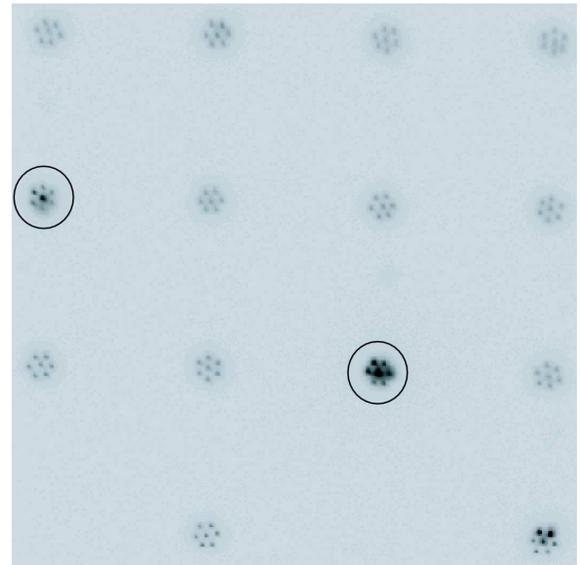}
  \end{center}
  \caption{Image of the guide-fibre bundles. There are 14 bundles each
of which consists of 7$\times$50$\mu$m diameter fibres. All fibres
are illuminated by sky emission and two bundles indicated
with large circles are aligned to guide stars and used for
auto-guiding.
}\label{fig:gfb}
\end{figure}

The guide-fibre bundle consists of 7 fibres with a core diameter 
of 50$\mu$m arranged in a hexagonal shape with 80$\mu$m spacing, 
which corresponds to 0$.\!\!^{\prime\prime}$96 at the focal plane. 
There are 14 guide-fibre bundles located at both edges of the field 
of view, marked with circles indicating the covering area 
of $\sim$3$^\prime$ in figure \ref{fig:fov}. The images of bright stars 
acquired by these guide-fibre bundles are shown in figure \ref{fig:gfb}. 

\begin{figure}[htbp]
  \begin{center}
    \FigureFile(75mm,75mm){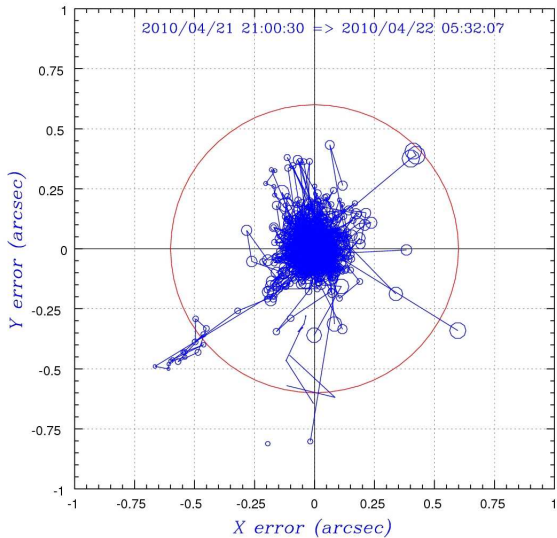}
  \end{center}
  \caption{Distribution of the position error during the auto-guide in 
the fixed coordinate (XY) to Echidna. The fibre diameter of 
1$.\!\!^{\prime\prime}$2 is indicated by the circle at the centre. 
Almost all the points located at the outside of the circle are sampled 
before auto-guiding is started. 
The size of the symbols corresponds to the averaged brightness of 
the guide stars. The median auto-guiding error is 
0$.\!\!^{\prime\prime}$067,  and typically less than 
0$.\!\!^{\prime\prime}$18. 
}\label{fig:pos1}
\end{figure}

The position error for the telescope pointing is calculated by the 
weighted average of the position offsets of these stars, 
which is typically less than 0$.\!\!^{\prime\prime}$18 under the typical sky 
condition (figure \ref{fig:pos1}). We can expect at least a few guide 
stars brighter than $R$=16.5 mag even in the high galactic-latitude area.

\begin{figure}[htbp]
  \begin{center}
    \FigureFile(75mm,75mm){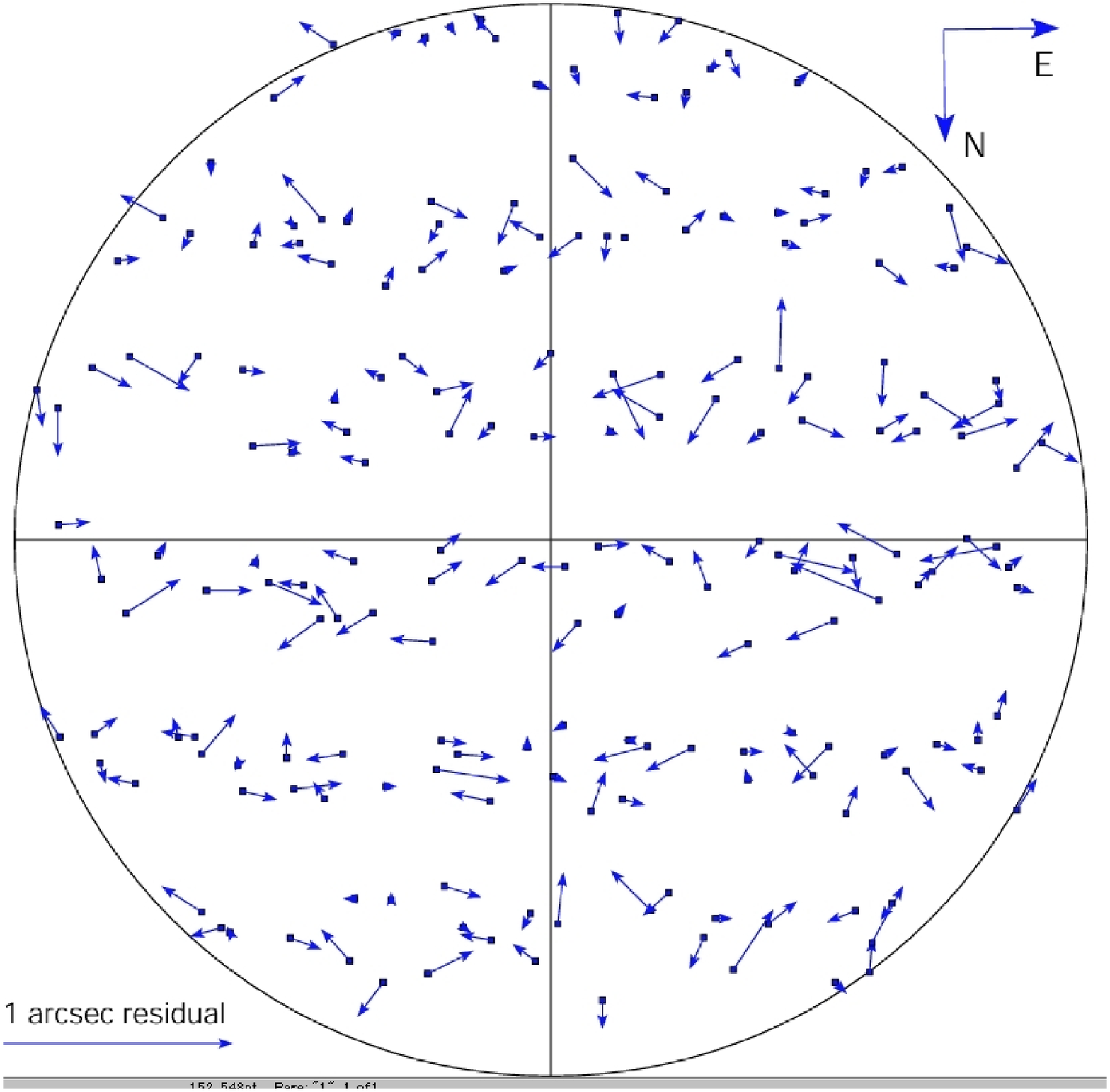}
  \end{center}
  \caption{Position offsets between the targets and science fibres
measured by the rastering observation in a open cluster field (NGC6633).
The big circle indicates the science field of view of FMOS and
each arrow indicates a offset measured with one science fibre.
The length of the arrows is exaggerated and the left bottom arrow
represents 1$.\!\!^{\prime\prime}$0 offset. 
The rms of the offsets is 0$.\!\!^{\prime\prime}$15. The
offsets are random and there is no systematic offsets due to
an incorrect distortion pattern modelling.
}\label{fig:map0}
\end{figure}

The position error for each fibre is estimated by ''rastering'' 
of the telescope in a grid pattern around the target position. 
The map of the observed flux of each object on all the grids of 
the raster pattern represents the position offset of the target 
from the science fibre. 
A maximum of 400 position offsets can be measured simultaneously 
by this raster sampling method, which determines the 
configuration accuracy of 
0$.\!\!^{\prime\prime}$15 (figure \ref{fig:map0}).

\subsection{Characteristics of the spectrographs}
First, we describe the results of the readout system on the detector. 
Then, we describe the thermal noise from the optics of spectrograph.

The typical exposure time of FMOS is expected to be $\sim$ 5-15 minutes to 
observe faint targets. Since very fast readout of the detector compared with 
the expected exposure time is not required, we operate the HAWAII-2 array 
with the four-channel readout mode, taking 17 seconds to make a single
read with IRS1, and taking 15 seconds with IRS2.
We use "ramp sampling mode" (iteration of a single read without
resetting the detector) in the usual scientific exposure to reduce 
the contribution of the read-noise of the electronics and also to enable the 
acquisition of bright objects among faint targets.

\begin{figure}[htbp]
  \begin{center}
   \FigureFile(70mm,70mm){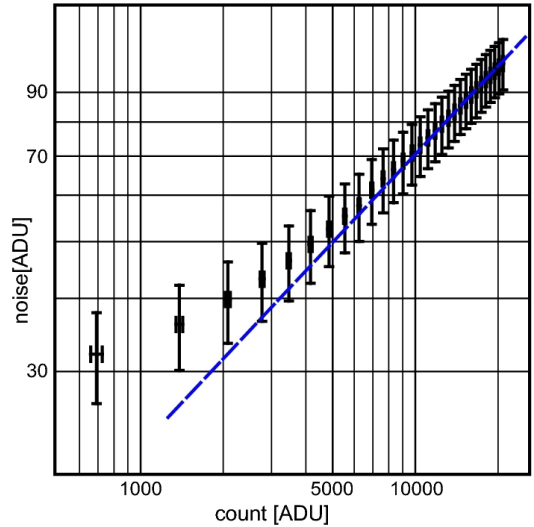}
  \end{center}
  \caption{
Conversion factor measurement. 
The broken line indicates the least square estimation 
between count and noise.
Measured Conversion factor is 2.0 [e$^-$/ADU]. 
The readout noise is 22 [e$^-$ rms] at IRS1.
The dashed line is a least squares fit to high count data.
}\label{fig:cvn}
\end{figure}

The conversion factor and the read-noise of the detector can be 
estimated from the count-noise diagram derived from many sets of 
ramp sampling data with a cover 
attached to the entrance window of the camera dewar. 
The conversion factor estimated for IRS1 from the diagram 
is $\sim 2.0$ electrons/ADU and the readout noise is 
$\sim 22$ electrons for a single read (figure \ref{fig:cvn}). 
The slope of the dashed line is 1/2, which is consistent with shot noise.

Since the maximum ADU count of 2$^{16}$=65536 corresponding 
to $\sim 1.3\times 10^5$ electrons is about half of 
the full electron capacity of the detector, 
the non-linearity of the detector is very small even 
when the count is high. The measured non-linearity 
is typically less than 1\% up to 40000 ADU at IRS1.

\begin{figure}[htbp]
  \begin{center}
    \FigureFile(70mm,110mm){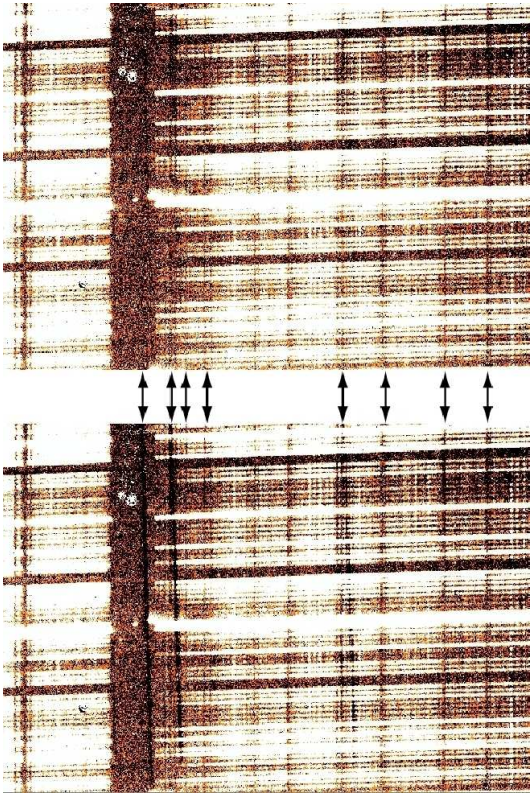}
  \end{center}
  \caption{Correction of crosstalk. 
The effect of the corrections is presented comparing the 
image before correction(top) with after(bottom). 
The positions of the crosstalk images are indicated by the arrows.
}\label{fig:crs}
\end{figure}

When very bright objects are included in the targets, we have to consider two 
effects on the image; crosstalk and latent images. The crosstalk is 
identified as ''negative'' images on the other quadrant of the detector 
where the counts are sampled at the same timing as the region with 
very high counts. The contribution of the 
crosstalk is about $-0.15\%$ of the count of the bright region, 
correctable in the later data reduction process (figure \ref{fig:crs}). 
The latent images are ''positive'' 
patterns remaining after an exposure including highly saturated 
region, caused by released charges into the conduction band 
of the detector array. It is difficult to correct the latent 
images because they depend on various parameters such as 
incident flux during the previous exposure, elapsed time, 
temperature, and so on.
It takes about 7 minutes until the latent image decays. 

\begin{figure}[htbp]
  \begin{center}
    \FigureFile(75mm,75mm){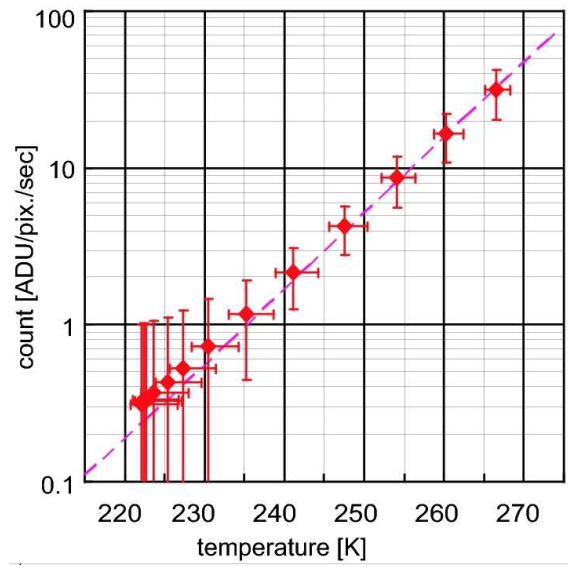}
  \end{center}
  \caption{A background flux as a function of temperature.
Points shows the measured background count[ADU/pixel/sec]. 
The dashed line is an expected thermal background flux.
}\label{fig:bft}
\end{figure}

The rejection capability of the thermal background of the IRS1 camera 
part was estimated from the correlation between the background 
count rate on the detector and the temperature of the refrigerator. 
In IRS1, we use a normal large refrigerator with pressurised dry air.
In IRS2, the cooling is performed with 
dry chilled air enclosure 
by a Polycold chiller and ducted into the enclosure.
Figure  \ref{fig:bft} shows the measured background flux for various 
temperatures of the refrigerator plotted with the expected correlation 
calculated from the transmission curve of the lens materials and 
the thermal-blocking filter multiplied by the quantum efficiency 
of the detector. 
The measured background flux is consistent with the expected flux 
indicating that the thermal-blocking filter works effectively.

\subsection{Total performance}
At first, a black body source with a temperature of 1095$^{\circ}C$ was 
attached to the Cassegrain focus to measure the system efficiency. 
\begin{figure*}[thbp]
 \begin{center}
  \FigureFile(67mm,67mm){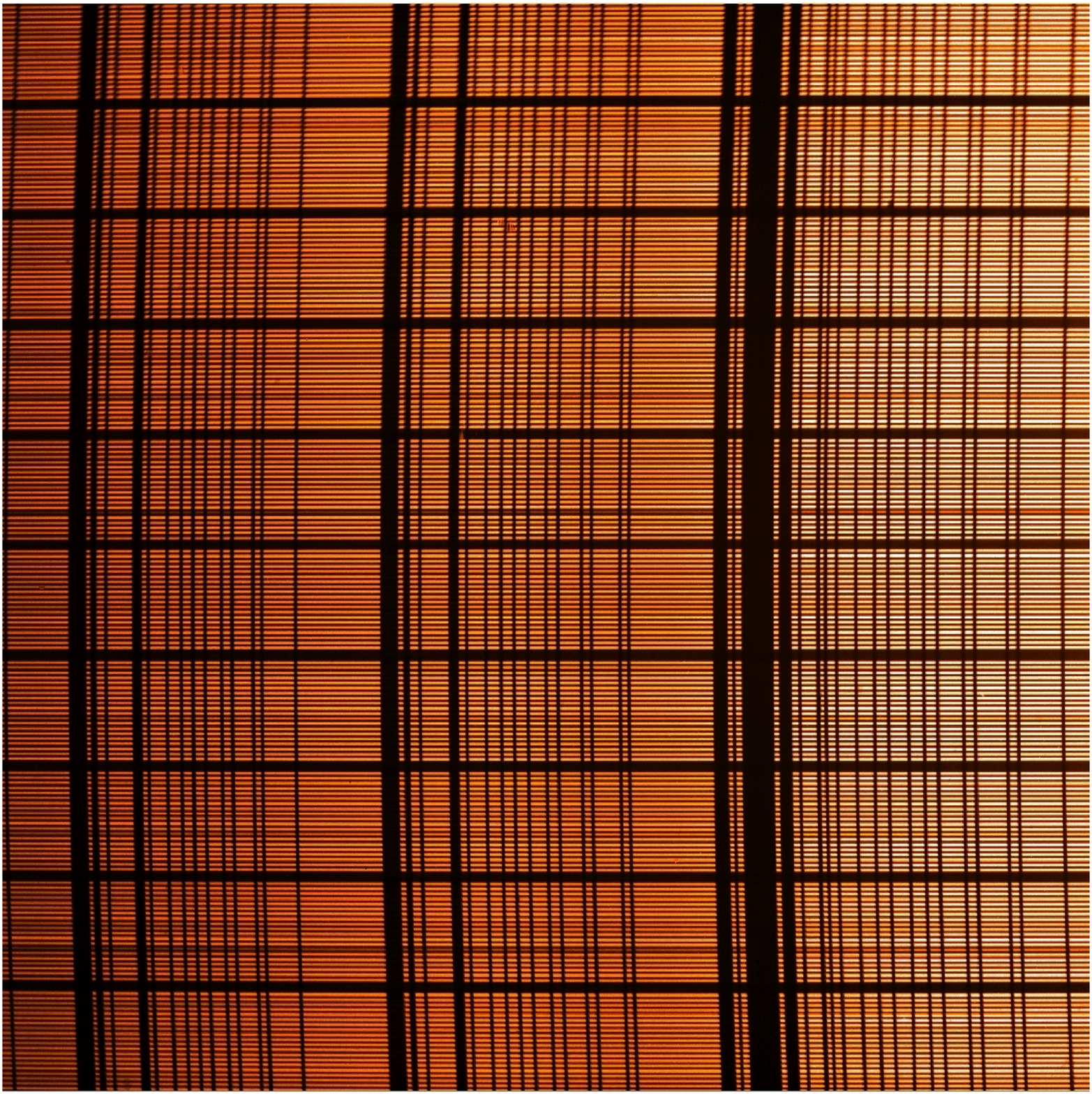}
\quad
  \FigureFile(67mm,67mm){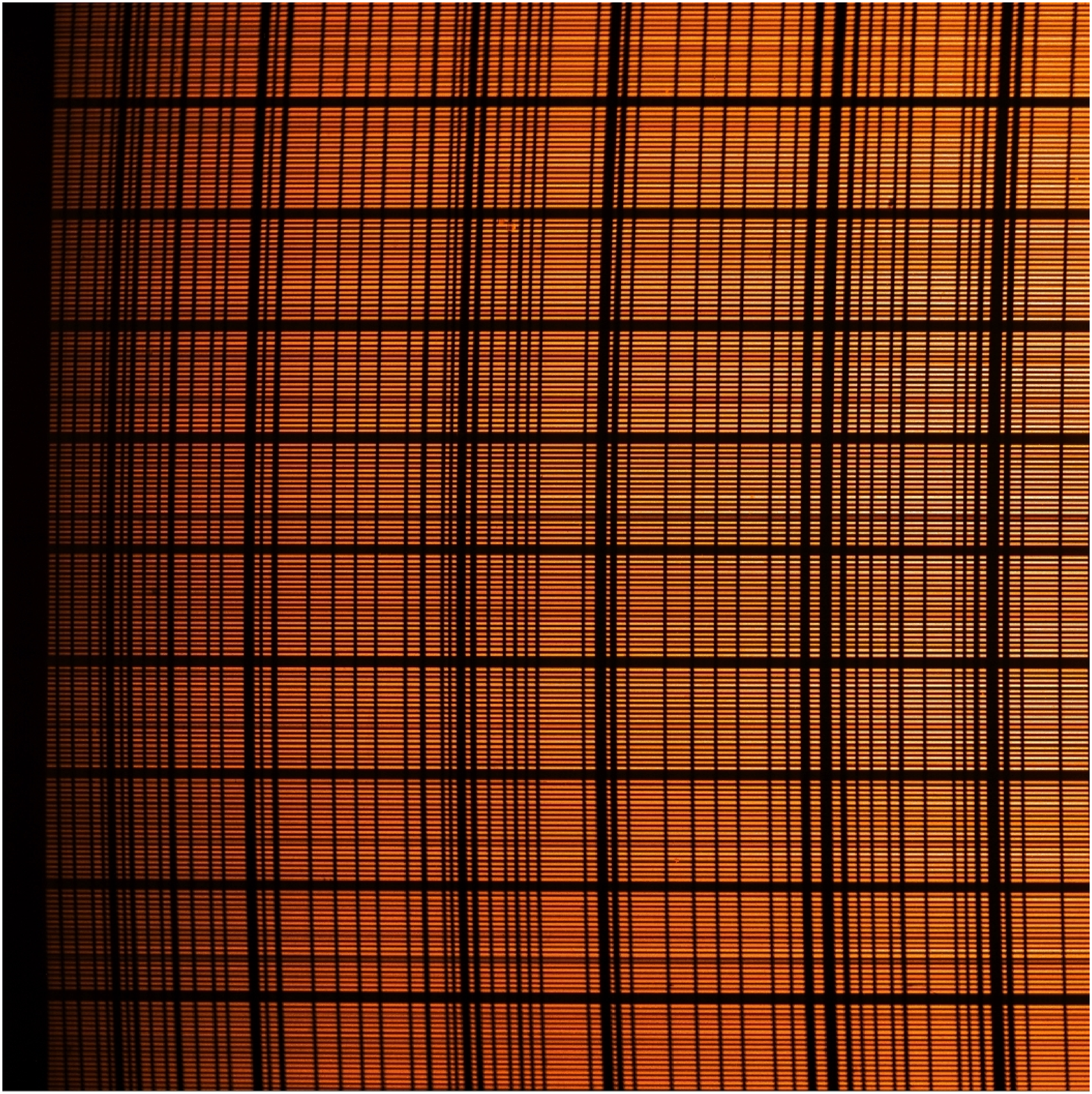}
 \end{center}
 \caption{The spectra of a blackbody emission 
with a temperature of 1095$^{\circ}C$ 
in the J-long(left) and H-short(right) bands. 
}\label{fig:cmh}
\end{figure*}
Figure  \ref{fig:cmh} shows the spectra of the blackbody source 
obtained by IRS1 in the $J$-long and $H$-short bands.
The typical FWHM of the vertical width of each spectrum 
is 5 pixels including the optical aberration of 2 pixels,
with a pitch of 10 pixels between the spectra (figure \ref{fig:flux}).

\begin{figure}[bhtp]
  \begin{center}
    \FigureFile(70mm,70mm){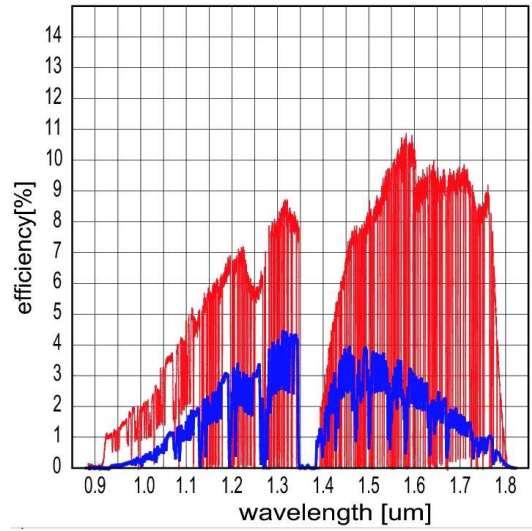}
  \end{center}
  \caption{Absolute efficiencies of the system from the prime 
focus corrector to the detector include the quantum efficiency 
of the detector with high- and low-resolution mode. 
Thin line is the typical(median) value of measured efficiency 
with high-resolution mode using a blackbody source at 1095 $^\circ$C, 
measured by the four exposures of different high-resolution modes. 
Thick line is the typical(median) value of measured efficiency 
with low-resolution mode, the throughput of the VPH grating 
is included. 
The maximum  efficiencies are about 3 \% larger than these graphs. 
}\label{fig:f21}
\end{figure}

Figure \ref{fig:f21} shows the measured efficiencies with 
low- and high-resolution mode in the full wavelength range.
Typical system efficiency for high-resolution mode 
is 6\% in $ J$ and 10\% in $ H$ from the prime focus 
corrector to the detector, almost consistent 
with the expected value at high-resolution mode. 
The decreases of the efficiency in the shorter wavelength side 
is due to the grating, which has the peak of 
the diffraction efficiency around 1.35 $\mu m$.
On the other hand, 
the decreases in the longer wavelength edge 
is due to the thermal cut-off filter in the camera dewar.
Furthermore, the gradual absorption feature from 1.3$\mu$m 
to 1.4$\mu$m is due to the attenuation of the fibre and other silica 
material used in almost all the lenses. 
The 1.35 - 1.40 $\mu$m part has to be blocked by the fibre slit.
Furthermore, these measured efficiencies have about 30 \% of 
the fibre-to-fibre variabilities. 
These are caused by grime on fibre edges as well as 
the throughput variety of the fibres 
including a misalignment of connectors.
We also found that the focal ratio degradation effect is not negligible, 
the factor of $\sim$0.8 estimated from the relative efficiency map 
for various positions of the black body source on the primary mirror cover. 

\begin{figure}[bhtp]
 \begin{center}
  \FigureFile(70mm,70mm){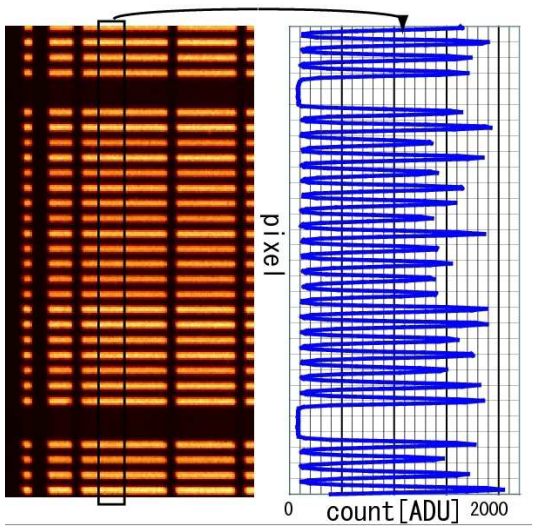}
 \end{center}
 \caption{Closeup view of the figure \ref{fig:cmh}(left).
 The right panel shows the count(horizontal axis) to pixels
(vertical axis, 10 pixel/lines) 
along the vertical line in the left figure.
}\label{fig:flux}
\end{figure}

We also measured the total efficiency with on-sky observation.
There is some difference between the on-sky observation 
and the blackbody test.
One part is the throughput before the corrector; the reflectivity of 
the primary mirror and the sky transmittance, 
the other part is the loss at the input surface of fibres; 
the position error of the fibre, 
a mis-match of the input light for the spine-tilt effect, 
a flux loss for the effect of seeing fluctuation, 
and the focal ratio degradation effect. 

Initially we measured the observational efficiency of the system 
using an image of a defocused bright star. 
The measured efficiency is 8\% in $H$ bands, almost consistent with 
the result of the blackbody test assuming the throughput. 
On the other hand, the total system efficiency
based on the observation of an open cluster is 7$\pm$3\% in $H$ bands, 
which is factor 0.9 lower than the measured value using the defocused 
bright star. 
Therefore, the loss at the input surface of fibres 
is not negligible during observations.

\begin{figure}[bthp]
 \begin{center}
  \FigureFile(80mm,60mm){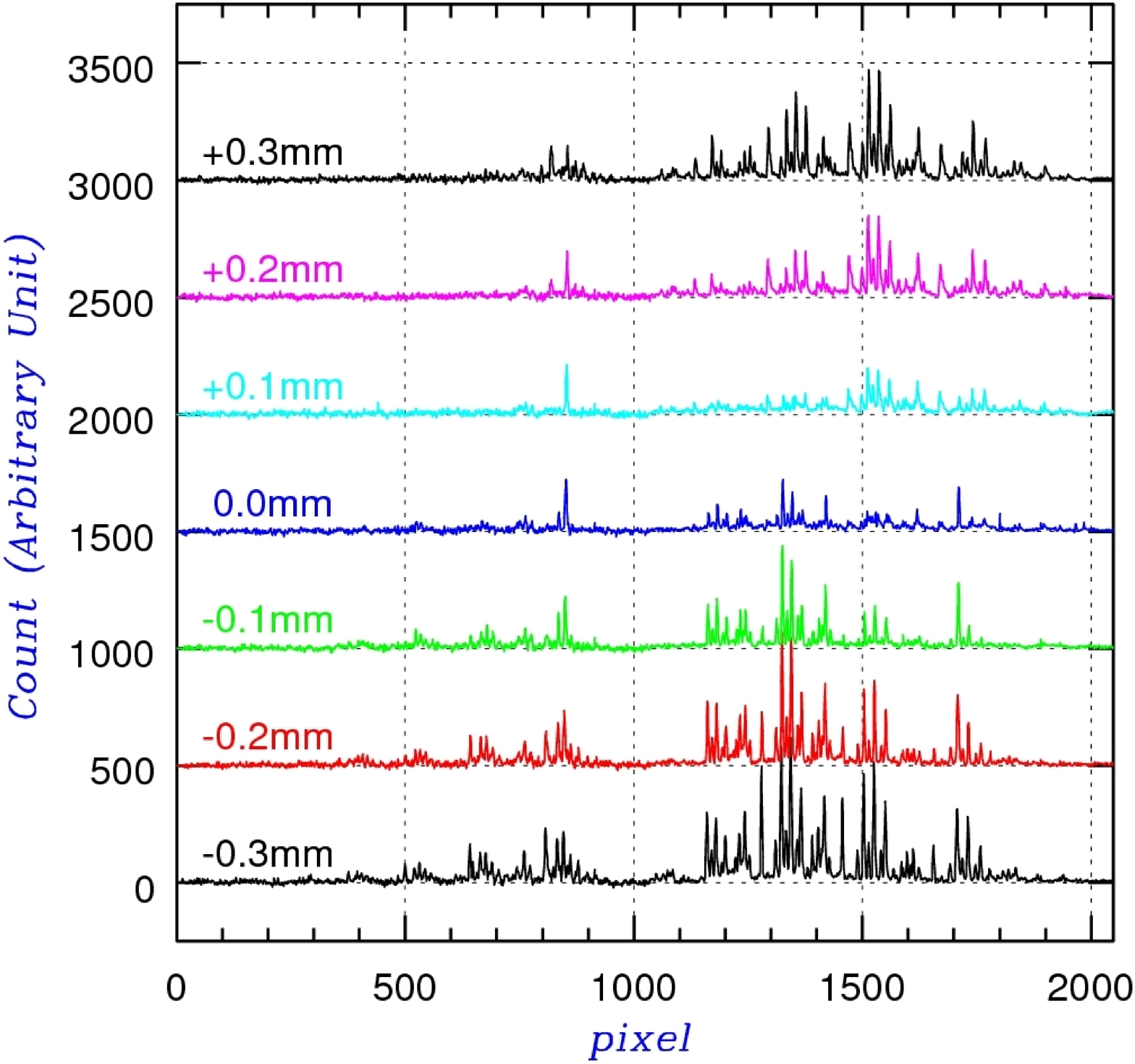}
 \end{center}
 \caption{
The variation of the airglow intensity for various offsets of the slit 
positions from +0.3mm to -0.3mm. The best position is $\sim$ +0.1mm.
}\label{fig:ohlm}
\end{figure}

The airglow rejection capability of the spectrograph can be tested by scanning 
the fibre slit along the dispersion direction slightly. Figure \ref{fig:ohlm}
shows the variation of the airglow intensity for various offsets of the slit 
positions, indicating that the mask configuration is not perfect yet. 
The expected rejection factor of the OH-airglow lines is $\sim 10$ 
after we make a little more improvement in the mask to get 
the maximum performance of the spectrograph.

\begin{figure}[bhtb]
 \begin{center}
  \FigureFile(80mm,130mm){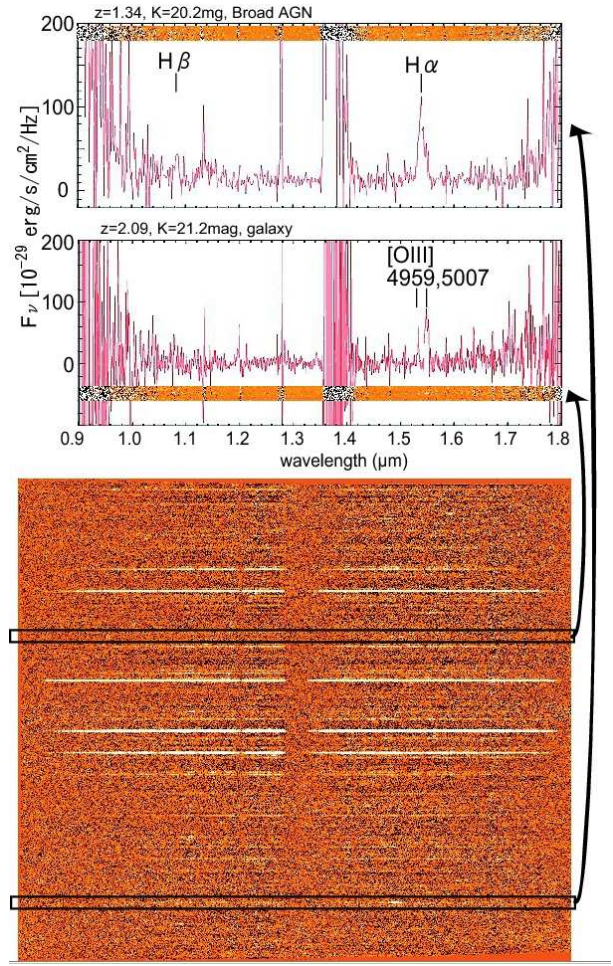}
 \end{center}
 \caption{
Two samples of extracted spectra of z=1.34 broad AGN (top) 
and z=2.09 galaxy (middle) 
from a reduced spectral image (bottom) in the SXDS field. 
The wavelength increases from the left (0.9$\mu$m) 
to the right(1.8$\mu$m). The bright four spectra are stars 
for position confirmation.
The emission lines is clearly detected H$\alpha$ 
and H$\beta$ (upper spectrum), and O[III] 
(bottom spectrum).
}\label{fig:sxdf1}
\end{figure}

Some scientific exposures have been carried out during the engineering 
observation. Figure \ref{fig:sxdf1} shows the reduced spectral image 
taken in the SXDF (Subaru/XMM-Newton Deep Survey) field 
from six sets of 15 minutes on- and off-source 
exposure (total 180 minutes exposure time) in the low-resolution mode. 
The magnitudes of the four bright stars in this image 
are around 17 mag. in $K$-band.
The centre gaps in the spectra correspond to the gap 
between the two mask mirrors where the fibre slit is located.

Although the sky and instrument conditions were not ideal, 
the estimated limiting 
magnitudes of continuum flux for an hour exposure 
with S/N$=$5 are $J=$20.1 mag 
and $H=$19.8 AB mag in the low-resolution modes. 
Furthermore, 
an emission line flux of $1\times10^{-16}$ [erg cm$^{-2}$ s$^{-1}$] 
is detected with S/N=5 by 1 hour integration both in J and H.

\section{Conclusions}
FMOS is a second-generation common-use instrument for the Subaru Telescope, 
having the capability to simultaneously acquire NIR spectra 
from up to 400 targets within a 30$^\prime$ field of view. 
After all the components were installed on the telescope 
in late 2007, the total performance was checked through various tests 
and engineering observations. The major results are as follows.
\begin{enumerate}
\item Sufficient configuration accuracy can be achieved with 
7 iterations using FPI; on average 98$\%$ of the fibres
reach target positions within 10$\mu$m at above zenith
distance of 60 degree. Echidna takes about 15 minutes 
to complete 7 iterations.
\item The configuration accuracy of fibres measured by the raster sampling 
method is less than 0$.\!\!^{\prime\prime}$15. The accuracy of auto-guiding 
using guide-fibre bundles is typically less than 0$.\!\!^{\prime\prime}$18.
\item The typical system efficiency of 10$\%$ from the prime focus
corrector to the detector is consistent with the expected value. 
The total system efficiency of 7$\%$ based on real observations is somewhat 
lower than expected. 
\item The rejection capability of the thermal background of IRS1 almost 
approaches the ideal value, indicating the refrigerator system and the 
thermal-blocking filter are working well. Although the blocking capability 
of the OH-airglow mask is currently sub-optimal, 
a blocking factor of $\sim 10$ 
is expected after a little more improvement in the mask.
\item Currently, the limiting magnitudes 
for a 1 hour exposure 
with S/N=5 are $J =$ 20.1mag and $H =$19.8 mag 
in the low-resolution modes. 
The emission line flux of $1\times10^{-16}$ [erg cm$^{-2}$ s$^{-1}$] 
is detected with S/N=5 by 1 hour integration both in J and H.
These values were measured from the results of the short observation in 
the SXDS field.
\end{enumerate}

Although IRS1 has almost been completed, it still has scope for improvement.
The blaze-wavelength of the mosaic grating, the cut-off wavelength 
of the thermal-blocking filter
and the deterioration of the total efficiency in low-resolution 
mode (due to a mis-match of the line density of the VPH grating) 
can all be improved.

If we use gratings having a 
shorter blaze-wavelength as well as the thermal-blocking filter with a shorter 
cut-off wavelength, the limiting magnitude in the $J$-band will be improved 
considerably.

\bigskip

\section*{Acknowledgment}
The FMOS project is managed and operated by the Kyoto university 
and National Astronomical Observatory of Japan. 
The present results were accomplished during engineering observation of
the Subaru telescope.
Tomonori Usuda and Daigo Tomono contributed importantly to the
accomplishment of the FMOS project. 
Kentaro Aoki and operational members support of these engineering observation. 
We wish to thank Tomio Kanzawa, Yoshitake Nabeshima, Kazuhito Namikawa, 
Satoshi Negishi, Masami Yutani, and James Ferreira 
for supporting out test observations at Subaru telescope.
The authors wish to thank Day crew members, Telescope group members, 
and Software members for valuable discussion and continuous 
support of this work. 
We also thank Scot Kleinman and Steve Colley for useful comments 
and suggestions.
We would like to express our thanks to the engineering staff of
Mitsubishi Co. for their fine operation of the telescope.
The UK component of the FMOS project was supported by grants from 
PPARC (now STFC), and we acknowledge helpful discussions with Ian Parry,
Sue Worswick, Derek Ives, Maggie Aderin, and Mattias Wallner.
Echidna fibre positioner is managed and developed 
by the Anglo-Australian Observatory,
and we acknowledge helpful discussions with all AAO's staffs. 
We also thank Hiroshi Karoji and Hiroshi Ohtani for initial contribution 
of this project.
This work was also supported by a Grant-in-Aid for Scientific
Research(A), Japan(17204016) and Scientific Research(B), Japan(17403003).

\end{document}